\documentclass[aps,apl,twocolumn,groupedaddress]{revtex4}

\input epsf
\usepackage{amsmath}

\begin{document}


\title{Hysteresis in superconducting short weak links and $\mu$-SQUIDs}

\author{Dibyendu Hazra$^{1}$, L\ae titia Pascal$^{2}$, Herv\'{e} Courtois$^{2}$, and Anjan K. Gupta$^1$}
\affiliation{$^1$Department of Physics, Indian Institute of Technology Kanpur, Kanpur 208016, India}
\affiliation{$^2$Institut N\'{e}el, CNRS and Universit\'{e} Joseph Fourier, 25 avenue des Martyrs, Grenoble, France. }
\date{\today}

\begin{abstract}
Thermal hysteresis in a micron-size Superconducting Quantum Interference Device ($\mu$-SQUID), with weak links as Josephson junctions, is an obstacle for improving its performance for magnetometery. Following the ``hot-spot" model of Skocpol et al. [J. Appl. Phys. {\bf 45}, 4054 (1974)] and by incorporating the temperature dependence of thermal conductivity of superconductor using a linear approximation, we find a much better agreement with the observed temperature dependence of the retrapping current in short superconducting Nb-based weak links and $\mu$-SQUIDs. In addition, using the temperature dependence of the critical current, we find that above a certain temperature hysteresis disappears. We analyze the current-voltage characteristics and the weak link temperature variation in both the hysteretic and non-hysteretic regimes. We also discuss the effect of the weak link geometry in order to widen the temperature range of hysteresis-free operation.
\end{abstract}


\maketitle
\section{Introduction}
A micron-size superconducting quantum interference device ($\mu$-SQUID) consists of two superconducting Dayem bridges or weak links (WL) \cite{likharev}, of dimension of the order of the superconducting coherence length, in parallel, forming a loop with area in the $\mu$m$^{2}$ range. A single WL behaves very much like a Josephson Junction \cite{likharev} with the supercurrent approximately given by $I = I_{c}\sin\theta$, where $I_{c}$ is the critical current and $\theta$ is the phase difference across the junction. When two such junctions are fabricated in parallel in a SQUID, interference between the two current branches gives an oscillatory behavior of the critical current $I_{c}$ with the external magnetic field \cite{tinkham-book}. The flux period is equal to the flux quanta $\Phi_{0}$ = 2.05 $\times$10$^{-15}$ T.m$^{2}$. This makes the SQUID a very sensitive device to measure magnetic field. While the flux sensitivity achieved by conventional SQUIDs is better than $10^{-7}\Phi_{0}/\sqrt{Hz}$, for a $\mu$-SQUID it has only been about $10^{-4}$$\Phi_{0}/\sqrt{Hz}$ \cite{wernsdorfer}. $\mu$-SQUIDs have been used to study the magnetization reversal \cite{wernsdorfer} of an isolated magnetic nano particle, the persistent current in phase-coherent rings \cite{rabaud} and also in scanning SQUID microscopy \cite{hasselbach}. An improved sensitivity of $\mu$-SQUIDs would be useful for probing ferromagnetic particles of smaller size or where the surface spins play an important role \cite{surf-spins}.

Other than the sensitivity, the hysteresis in $\mu$-SQUIDs current-voltage (I-V) characteristic (see e.g. Ref. \onlinecite{hasselbach}) is a major hurdle as it (1) increases the measurement time, (2) complicates the measurement electronics, (3) changes the temperature of the sample placed in close proximity with the $\mu$-SQUID. Thus it is important to understand this hysteresis and devise ways of eliminating it. Hysteresis in the current-voltage characteristic is a very common phenomena for many superconducting nano-structured systems, especially WLs. It includes conventional Superconductor-Insulator-Superconductor (S-I-S) Josephson junctions \cite{tinkham-book}, Superconductor-Normal metal-Superconductor (S-N-S) junctions \cite{angers,crosser}, superconducting nano-wires \cite{rogachev} and superconducting $\mu$-bridges \cite{fulton,tinkham-jap,song}. When the current is ramped up from zero across such junctions, the system suddenly switches to a non-zero voltage state at the critical current $I_{c}$. After switching, when the current is ramped down, the system comes back to a zero-voltage state at a particular current, called the retrapping current $I_{r}$. At very low temperature, the retrapping current can be smaller than the critical current: $I_r < I_{c}$. This defines an hysteretic I-V curve.

A number of models have been proposed in the last few decades to understand the hysteresis in superconducting WLs. The resistively and capacitively shunted junction (RCSJ) \cite{tinkham-book} model predicts the I-V curve for a conventional S-I-S junction very well. In this case, the capacitance across the junction is responsible for the hysteresis. But for lateral junctions (either S-N-S junctions or constrictions), the geometrical capacitance is too small to explain hysteresis. Hence an alternative theory of an effective capacitance $C_{eff}$ was proposed \cite{song}, where one equates the charge relaxation time $R_{n}C_{eff}$ with the Cooper pair relaxation time $h/\Delta$. Here $R_{n}$ is the normal resistance and $\Delta$ is the superconducting gap parameter. The same method was recently extended to S-N-S junctions \cite{angers} by equating $R_{n}C_{eff}$ with the diffusion time of Andreev pairs. Though these methods reproduce some of the features of the I-V curves, no justification behind the origin of an effective capacitance has been found.

Recently, Courtois et al. \cite{courtois} have unambiguously shown, by directly measuring the electronic temperature, that heating is responsible for hysteresis in S-N-S junctions. According to the ``hot-spot" model of Skocpol et al. \cite{tinkham-jap}, the heat generated in the resistive region of the WL raises locally its temperature above the critical temperature $T_c$. The temperature goes down to the bath temperature as one moves away from the hot spot. This gives rise to a normal metal-superconductor interface along the surface defined by $T=T_c$. The interface location is self-consistently determined by the heat generated and the coupling to the thermal bath. It was found that below a certain current, identified as the retrapping current, this interface becomes unsustainable and the WL turns fully superconducting. For a short WL, the hot spot may spread beyond the WL and into the electrodes. This ``hot-spot" model reproduced most of the features of the I-V characteristics of superconducting WLs. It also predicted a $\sqrt{1-T/T_c}$ dependence of $I_r$ on $T$; however the latter was not experimentally verified. Further, this model ignored the temperature dependence of the thermal conductivity of superconductor. Incorporating an approximate form for this temperature dependence, Tinkham et al. described the I-V characteristics of free standing superconducting nano-wires \cite{tinkham-prb}. In this case, the N-S interface occurs inside the long nano-wire, making the problem one-dimensional. Again, this work did not include the temperature dependence of the hysteresis in the I-V characteristics.

In this paper, we describe an effective one-dimensional thermal model to find out the temperature profile near a short WL connected to wide electrodes. We calculate the I-V characteristics as well as the (bath) temperature dependence of the retrapping current. Our model predicts how the normal-superconducting (N-S) interface position varies with various parameters like temperature, current, and geometry. We also discuss the detailed temperature profile and how it changes with the bias current. Using the temperature dependence of the critical current near $T_c$, we find that above a certain temperature $T_h$, hysteresis disappears. The effect of the WL geometrical parameters on $T_h$ is discussed quantitatively. As the same model is directly relevant to $\mu$-SQUIDs, we test it on several such samples. Our model fits our data very well. Finally, we discuss how the non-hysteretic regime can be achieved over a wider temperature range, followed by conclusions.

\section{Thermal model of hysteresis}

\begin{figure}[tbp]
\centerline{\epsfxsize = 1.8 in \epsfbox{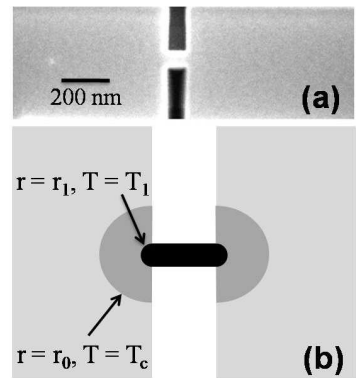}}
\caption{(a) Scanning electron micrograph image of a WL. (b) Sketch of the sample geometry with the three regions discussed in the text.}
\label{fig:wl}
\end{figure}

Following the ``hot-spot" model \cite{tinkham-jap}, we consider a single WL connected to two extended electrodes, as shown in \mbox{Fig.} \ref{fig:wl}a, and investigate the temperature distribution around it in the resistive state. We assume a local quasi-equilibrium condition so that a local temperature can be defined at each point of the sample. The length and width of the WLs under study are in 50-200 nm range, \mbox{i.e.} of the order of the coherence length of bulk Nb ($\xi_{Nb}\approx$ 39 nm). In this range, a WL behaves very much like a Josephson Junction \cite{likharev}. Since the WL size is very small, we assume that in the resistive state the entire WL region stays at a uniform temperature. In reality, the WL will have certain spatial temperature variation, but what will matter here is the heat evacuated out of the WL. This assumption will not affect our conclusions as long as the WL temperature is not so large as to affect its resistance. We also assume the pads connecting to the WLs to be much wider than the length scale over which the temperature relaxes to the bath temperature.

At a given bias current, we can divide the device into three regions as shown in \mbox{Fig.} \ref{fig:wl}b: 1) the narrow WL at a uniform temperature $T_1$ consisting of the rectangular (width $w$ and length $l$) area in the center and terminating into a semicircle of radius $r_1=w/2$ at each end, 2) a normal state ($T  > T_{c}$) semicircular region in each electrode with $r_{0} > r > w/2$, and 3) a superconducting ($T < T_{c}$) region for $r > r_0$. We have assumed a rounded WL of radius $r_1 = w/2$ to avoid logarithmic divergence in the resistance calculation. This approach is unlike the hot spot model for a long WL \cite{tinkham-jap}, where the hot-spot develops near the center of the WL. Beyond the WL, we assume that the heat conducts away radially in the bulk of the film. Thus the temperature also decreases radially inside the two electrodes, reaching the bath temperature $T_b$ far away from the WL. This enables us to use an effective one-dimensional model for finding the spatial variation of temperature. The source of heat is the resistive dissipation in the normal region, which extends up to a radial distance $r_0$ in each electrode, thus defining a N-S interface with $T=T_c$ between normal and superconducting regions.

In the resistive state, the Joule heat near the WL region is removed in two ways: conduction within each electrode and surface heat flow from film's bottom surface to the substrate. The latter is assumed to be proportional to the temperature difference between the film and the substrate. This approximation has been used extensively \cite{tinkham-jap,peroz,lin-heat-rmp}; we will discuss its validity later. We neglect the heat loss from the top surface of the film as we operate in a vacuum cryostat. We also assume that the entire substrate stays at the bath temperature $T_{b}$. Thus, the general heat flow equation can be written as:
\begin{eqnarray}
\nonumber -\kappa \frac{dT}{dr}2\pi rd + \displaystyle\int^r_{r_1} \alpha (T-T_{b})2\pi r \,dr + \alpha (T_1-T_{b})A\\
= I^{2}R(r).
\label{eq:hf1}
\end{eqnarray}
Here $\kappa$ is the thermal conductivity, $d$ is the thickness of the film, $A$ is the area of the WL region given by $A=(wl+\pi {r_1}^2)$, $I$ is the current. The surface heat loss coefficient $\alpha$ is expressed in W/m$^{2}$K. $R(r)$ is the resistance of the electrode within a radius $r$ including the resistance of the WL. Assuming a radial and isotropic current flow in the electrodes giving rise to circular equipotential lines, the resistance $R(r)$ is given by:
\begin{eqnarray}
R(r) = R_0+R_c\ln(\frac{r}{r_{1}}),
\label{eq:resistance}
\end{eqnarray}
where $\pi R_{c}/2 = \rho_{N}/d$ is the square resistance of the film, with $\rho_N$ as the normal state resistivity. Thus $R_c$ defines a characteristic resistance of the film. The resistance $R_0$ of the WL is given by $R_0=(l+2r_1)\rho_N/wd$.

Since we consider low temperatures, we ignore the phonon's contribution to the thermal conductivity. In the normal state, the electronic part of the thermal conductivity $\kappa_n$ can be found using the Wiedemann-Franz law: $\kappa_n = L_0T/\rho_N$, where $\kappa_n$ is the thermal conductivity in the normal state and $L_{0}$ is the Lorentz number. In the superconducting state, as the number of quasi-particles decreases significantly, one expects the thermal conductivity $\kappa_s$ to be much less. At very low temperature, since only very few quasi-particles are left to carry thermal energy, $\kappa_{s}$ can be exponentially small. We use here a linear approximation, $\kappa_s/\kappa_n = T/T_c$, which gives $\kappa_s = \kappa_n$ at $T=T_c$ as expected. From the theory \cite{abrikosov}, this linear approximation is well justified near $T_{c}$. The same approximation was also used by Tinkham et al. \cite{tinkham-prb}.

Using the above expressions for $\kappa$ and $R(r)$ and differentiating \mbox{Eq.} \ref{eq:hf1} with respect to $r$, one gets:
\begin{eqnarray}
\nonumber \frac{1}{r}\frac{d}{dr}\left [rT\frac{dT}{dr}\right]-\frac{\rho_{N}\alpha}{L_{0}d}(T-T_{b}) = - \left(\frac{I\rho_n}{\pi d }\right)^{2}\frac{1}{L_{0}r^{2}},\\
(r_1 \leq r < r_0)
\label{eq:hf2}\\
\frac{1}{r}\frac{d}{dr}\left[rT^2\frac{dT}{dr}\right]-\frac{\rho_{N}\alpha T_{c}}{L_{0}d}(T-T_{b}) = 0 \hspace{1.5em}  (r > r_0).
\label{eq:hf3}
\end{eqnarray}
The boundary conditions are: 1) at $r=r_1$, $T= T_{1}$ and \mbox{Eq.} \ref{eq:hf1} gives $-\kappa(T_1) \frac{dT}{dr}2 \pi r_1 d + \alpha (T_1-T_{b})A = I^{2}R_0$, 2) at $r=r_{0}$, $T= T_{c}$, $T$ and $\frac{dT}{dr}$ are continuous and 3) for $r \rightarrow \infty$, $T = T_{b}$. The radius $r_0$ and WL temperature $T_1$ have to be found self-consistently using these boundary conditions.

An inspection of the above two equations gives us a length scale,
\begin{eqnarray}
\eta = \sqrt{\frac{L_{0}T_{c}d}{\alpha \rho_{N}}} = \sqrt{\frac{2L_{0}T_{c}}{\pi \alpha R_{c}}}
\end{eqnarray}
and a current scale
\begin{eqnarray}
I_{0} = \frac{\pi d T_{c}}{\rho_{N}}\sqrt{L_{0}} = \frac{2T_{c}}{R_{c}}\sqrt{L_{0}} = \frac{\pi\alpha}{\sqrt{L_0}}\eta^2.
\end{eqnarray}
Here $I_0$ would determine the scale of the retrapping current $I_r$, while $\eta$ would determine the length scale of temperature variation. For WLs based on a Nb film deposited on a Si substrate, one typically uses a thickness of 20 to 150 nm. Depending upon the detailed preparation method, some typical parameters would be $\rho_N$ = 15-50 $\mu\Omega$ cm, $T_c$ = 6-9 K and $\alpha$ = 1-3 W/cm$^2$ K \cite{peroz}. Using $L_0$ = 2.44$\times$10$^{-8}$ W.$\Omega$/K$^2$, we get $\eta\sim$1-3 $\mu$m and $I_0 \sim$ 0.5-2 mA.

\mbox{Eq.} \ref{eq:hf2} and \ref{eq:hf3} can be written in terms of the dimensionless variables $x$= $r/\eta$, $t = T/T_{c}$, $i = I/I_0$, $x_1$ = $r_1/\eta$, $x_0$ = $r_0/\eta$ and $t_b = T_b/T_c$ as follows:
\begin{eqnarray}
\frac{1}{x}\frac{d}{dx}\left[xt\frac{dt}{dx}\right] - (t-t_{b}) = - \frac{i^2}{x^2} \hspace{1.5em} (x_1 \leq x < x_0),
\label{eq:hf4}\\
\frac{1}{x}\frac{d}{dx}\left[xt^2\frac{dt}{dx}\right] - (t-t_{b}) = 0 \hspace{1.5em} (x > x_{0}).
\label{eq:hf5}
\end{eqnarray}
In terms of the reduced variables, the boundary conditions become: 1) at $x=x_1$, -$x_1t_1\frac{dt}{dx}+(t_1-t_b)\frac{A}{2\pi\eta^2}=\frac{\pi d R_0}{2\rho_N}i^2$, 2) at $x = x_0$, $t=1$ and $dt/dx$ is continuous, and 3) for $ x \rightarrow \infty$, $t = t_{b}$. For short weak links, using $\frac{A}{2\pi\eta^2}<<1$, the first boundary condition becomes $t_1\frac{dt}{dx}=-\beta i^2/x_1$ with $\beta = \frac{\pi R_0 d}{2\rho_N}$ = $\frac{R_{0}}{R_{c}}$ $\cong$ $\frac{\pi}{2}$$(1+\frac{\ell}{w})$.

Eq. \ref{eq:hf4} and \ref{eq:hf5} are second order and non-linear differential equations that can be solved only numerically. We wish to go beyond the approximation of $\kappa$ being independent of temperature, which would give solutions in terms of modified Bessel functions as discussed by Skocpol et al.\cite{tinkham-jap}. We choose to simplify the above equations by substituting $y_1=t^2$ and $y_2=t^3$ in \mbox{Eq.} \ref{eq:hf4} and \ref{eq:hf5}, respectively. $y_1$ and $y_2$ then satisfy:
\begin{eqnarray}
\frac{d}{dx}\left[x\frac{dy_1}{dx}\right] = - \frac{2i^2}{x}+2(\sqrt{y_1}-t_{b})x \hspace{1.5em} (x_1 \leq x < x_0),
\label{eq:hf6}\\
\frac{d}{dx}\left[x\frac{dy_2}{dx}\right]= 3(y_2^{1/3}-t_{b})x \hspace{1.5em} (x > x_{0}).
\label{eq:hf7}
\end{eqnarray}

\begin{figure}[tbp]
\centerline{\epsfxsize = 2.15 in \epsfbox{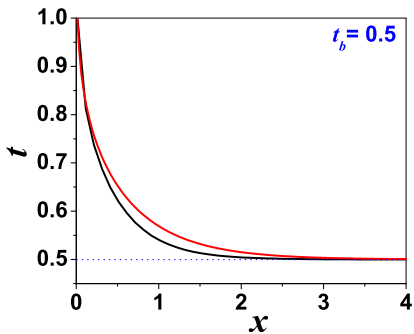}}
\caption{Comparison between exact (\mbox{Eq.} \ref{eq:hf7a}, black) and approximate (\mbox{Eq.} \ref{eq:hf8}, red) solution for the reduced temperature profile for $x>x_{0}$. The respective calculated slopes at the origin are - 5.1 and - 4.80. Parameters are $t_{b}=0.5$ and $x_{0}=0.015$.}
\label{fig:diff_eqn}
\end{figure}

Let us first consider the superconducting region ($x>x_0$) described by \mbox{Eq.} \ref{eq:hf7}. In this equation, $y_2$ varies between $t_b^3$ and 1. For this range of $y_2$, we linearly approximate the $(y_2^{1/3}-t_{b})$ term as:
\begin{eqnarray}
y_2^{1/3}-t_{b} \approx \frac{y_2-t_b^3}{1+t_b+t_b^2},
\label{eq:hf7a}
\end{eqnarray}
so as to keep the end points of ($y_2^{1/3}-t_{b}$), \mbox{i.e.} 0 at $t=t_b$ and $(1-t_b)$ at $t=1$, fixed. This approximation becomes more and more accurate as $t_b$ approaches 1, \mbox{i.e.} the bath temperature $T_b$ approaches the critical temperature $T_c$. \mbox{Eq.} \ref{eq:hf7} then reduces to the modified Bessel equation $s^2\xi''+s\xi'-s^2\xi=0$, where $\xi=y_2-t_b^3$ and $s=\lambda x$ with $\lambda =\sqrt{3/(1+t_b+t_b^2)}$. With the boundary condition $t=t_b$ (\mbox{i.e.} $\xi=0$) at $x\rightarrow \infty$, the only acceptable solution is $\xi=CK_0(s)$, where $K_0$ is the modified Bessel function of second kind and zero degree. Using the boundary condition $t=1$ at $x=x_0$, we get the final solution for $x>x_0$ as:
\begin{eqnarray}
t^3=t_b^3+\frac{1-t_b^3}{K_0(\lambda x_0)}K_0(\lambda x).
\label{eq:hf8}
\end{eqnarray}
\mbox{Fig.} \ref{fig:diff_eqn} shows for comparison the numerical solution of the non-linear \mbox{Eq.} \ref{eq:hf7} and the corresponding solution to the linearized \mbox{Eq.} \ref{eq:hf8} for $t_b=0.5$ and $x_0=0.015$. The overall shapes of the curves are similar, justifying our approximation.

Let us now consider the normal region ($x<x_0$). \mbox{Eq.} \ref{eq:hf6} is difficult to linearize as $y_{1}$ varies between $t_b$ and $t_1$, and $t_1$ is not known before-hand. The nature of boundary conditions does not allow a simple numerical solution. We make the approximation to neglect the surface loss term, \mbox{i.e.} the $(\sqrt{y_1}-t_b)$ term in \mbox{Eq.} \ref{eq:hf6}. This is justified for finding the retrapping current $I_r$ in the regime $x_0\gtrsim\ x_1$, in which case the heat loss to the substrate is not significant as compared to the heat conducted out. With this approximation, there is an analytical solution: $y_1=-i^2\left[(\ln x+C_1)^2)+C_2\right]$. Here $C_1$ and $C_2$ are constants to be found from the boundary conditions: $y_1=t^2=1$ at $x=x_0$ and $dy_2/dx=-2\beta i^2/x_1$. Finally, we get for $x_1<x<x_0$:
\begin{eqnarray}
t^2=1-i^2[(\ln\frac{x}{x_1}+\beta)^2-(\ln\frac{x_0}{x_1}+\beta)^2].
\label{eq:hf9}
\end{eqnarray}
This relation gives the temperature profile for $x<x_0$ and determines the WL temperature $t_1=t(x_1)$ in terms of $x_0$. To find $x_0$, we have to use the continuity of $dt/dx$ at $x_0$ using solutions given by \mbox{Eq.} \ref{eq:hf8} and \ref{eq:hf9}. This gives the following transcendental equation for $x_0$:
\begin{eqnarray}
i^2=\frac{\lambda x_0 (1-t_b^3)K_1(\lambda x_0)}{3[\ln(x_0/x_1)+\beta]K_0(\lambda x_0)}.
\label{eq:minima}
\end{eqnarray}

\begin{figure}[tbp]
\centerline{\epsfxsize = 2.5 in \epsfbox{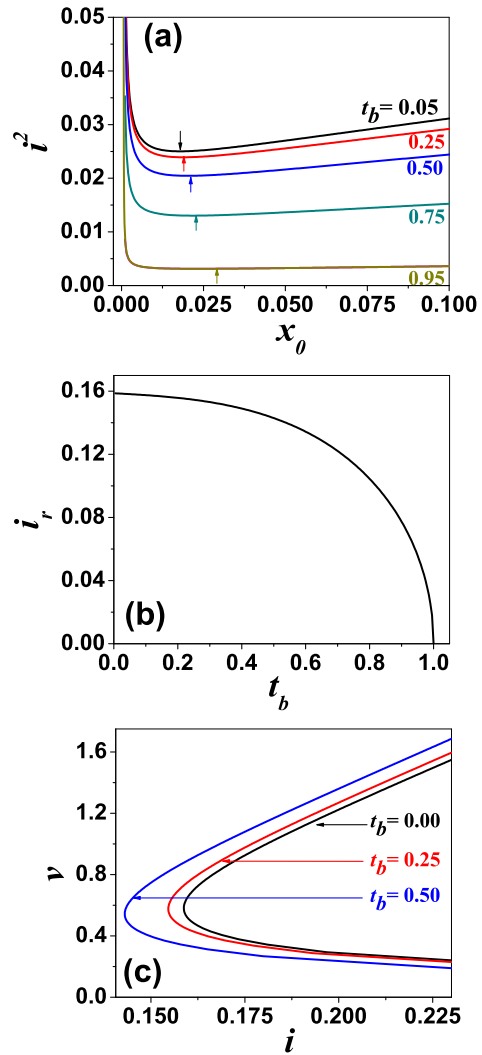}}
\caption{(a) Plot of \mbox{Eq.} \ref{eq:minima} right-hand side as a function of $x_{0}$ at different bath temperatures. The minima shown by arrows define the retrapping current $i_{r}$. (b) Variation of the retrapping current as a function of the bath temperature. (c) I-V characteristics near $i_{r}$ at three different bath temperatures as indicated in the figure. The current and the voltage axes are normalized with respect to $I_{0}$ and $I_{0}R_{c}$, respectively. All the curves are plotted for $\beta = 3.5$ and $x_{1} = 0.015$.}
\label{fig:mini}
\end{figure}

As shown in \mbox{Fig.} \ref{fig:mini}a, the right hand side of above \mbox{Eq.} \ref{eq:minima} features a minima in current $i$ as a function of $x_{0}$. This means that below this current, the Joule heat is not sufficient to uphold a normal metal-superconductor (N-S) interface with $T = T_{c}$. This current is thus identified as the retrapping current $i_{r}$ = $I_{r}$/$I_{0}$. \mbox{Fig.} \ref{fig:mini}b shows that it decreases with increasing bath temperature, whereas the related $x_{0}$ increases. At high temperature, a regime where the retrapping current exceeds the critical current ($i_r > i_c$) can be reached. In this case, the WL is resistive while its temperature stays below $T_c$. Only if the bias current becomes larger than $i_r$, does a N-S interface with $T=T_c$ appear at $r_0$ ($>r_1$). We will discuss this point in more detail later.

\mbox{Eq.} \ref{eq:minima} provides the relation between the current bias $I$ and the N-S interface position $x_{0}$. One can then calculate the resistance $R(x_0)$ using \mbox{Eq.} \ref{eq:resistance}. The related current-voltage characteristic $V= IR(x_{0})$ is plotted in \mbox{Fig.} \ref{fig:mini}c for different bath temperatures. At low voltage, a negative differential resistance branch appears, meaning that, in this regime, for a given current, the voltage can have two distinct values. Since this branch is unstable under current biasing \cite{ridley}, only the higher voltage is accessible. But if one performs voltage-biased measurements, then one can access the negative differential resistance branch as well, as was observed by Skocpol et al. \cite{tinkham-jap} in micro-bridges and Steinbach et al. \cite{steinbach} in Josephson junction.

\begin{figure}[tbp]
\centerline{\epsfxsize = 2.5 in \epsfbox{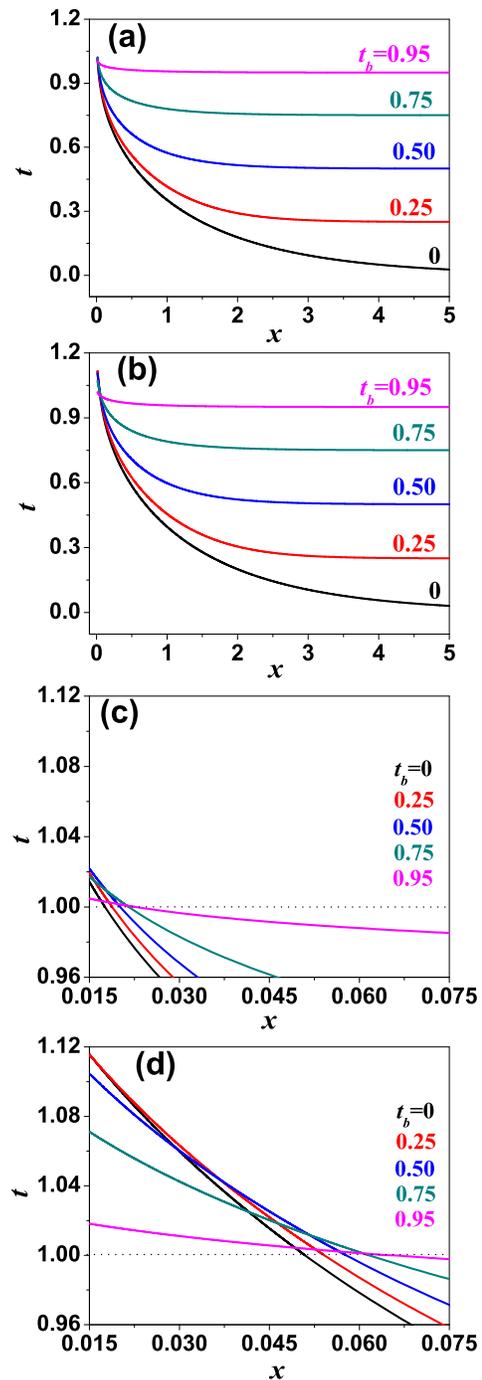}}
\caption{Temperature (normalized) evolution with radial distance (normalized) in a WL biased at its retrapping current at different bath temperatures indicated in the figures. The parameters are $x_1=0.015$ and (a) $\beta=3.5$, (b) $\beta=1.5$. (c) and (d) are magnifications near the normal region, corresponding to (a) and (b) respectively. The intersection with the dotted lines at $t=1$ indicate the N-S interface position.}
\label{fig:temp_dist}
\end{figure}

For illustration, let us now consider a WL biased at its retrapping current. \mbox{Fig.} \ref{fig:temp_dist} shows the radial temperature distribution for different bath temperatures and for some typical values of $\beta$ and $x_{1}$. Expectedly, at large distance, the temperature asymptotically decreases to the bath temperature. The temperature profile close to the WL exhibits a large temperature gradient as compared to the superconducting region, see \mbox{Fig.} \ref{fig:temp_dist}c, d. The intercepts of the different curves with the dotted horizontal lines representing $t=1$ indicate the location of the N-S interfaces. The temperature values at $x=x_1$ (= 0.015 here) indicate the WL temperature.

\begin{figure}[tbp]
\centerline{\epsfxsize = 3.5 in \epsfbox{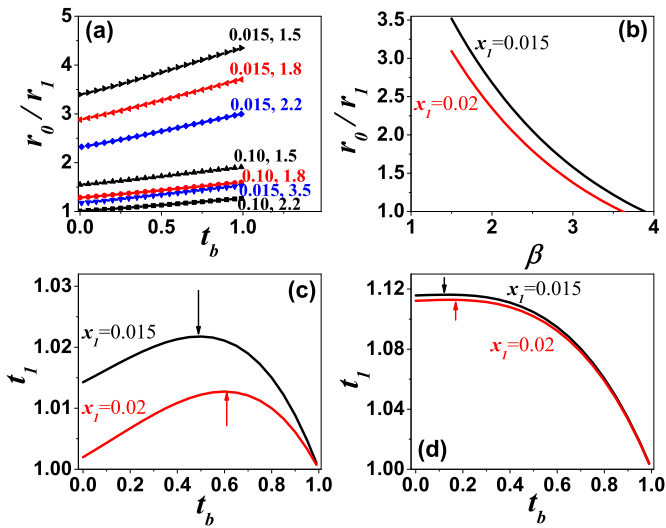}}
\caption{Variation of the N-S interface position $r_{0}$ in units of $r_{1}$ at the retrapping current as a function of: (a) bath temperature $t_b$ for different values of $x_{1}$ and $\beta$, (b) parameter $\beta$, at $t_{b}=0.15$ for different values of $x_{1}$. (c) and (d) show the variation of the WL temperature at retrapping as a function of bath temperature for (c) $\beta= 3.5$ and (d) $\beta=1.5$ for $x_{1}$ = 0.015 and 0.02. The arrows show the maxima.}
\label{fig:x0_tb}
\end{figure}

Still at retrapping, \mbox{Fig.} \ref{fig:x0_tb}a,b shows the variation of the N-S interface position $r_{0}$ in units of $r_{1}$ as a function of the bath temperature for different $x_{1}$ and $\beta$ values and as a function of $\beta$ at a fixed bath temperature for different $x_{1}$ values. For large values of $\beta$, \mbox{i.e.} for long WLs, the N-S interface is closer to the WL. \mbox{Fig.} \ref{fig:x0_tb}c and d show the temperature of the WL as a function of the bath temperature for a combination of $\beta$ and $x_{1}$ values. We observe a non-monotonic behavior, which is due to the increase of the thermal conductivity with increasing temperature. The contrast between, on one hand, the monotonic evolution of the current $i_r$ and the N-S interface position $x_{0}$ with $t_b$ and, on the other hand, the non-monotonic evolution of the WL temperature at retrapping indicates that it is the size of the normal region and not its local temperature that governs the amplitude of the retrapping current.

\section{Transport experiments on weak link $\mu$-SQUIDs}

\begin{figure}[tbp]
\centerline{\epsfxsize = 2.15 in \epsfbox{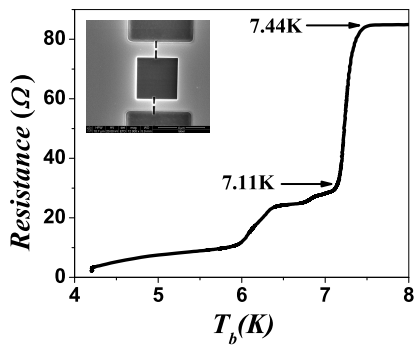}}
\caption{Resistance vs temperature curve for Sample 2 down to 4.2 K at a bias current of 0.1 mA. Inset shows the SEM image of a typical $\mu$-SQUID with a loop area 3.5 $\times$ 3.5 $\mu$m$^2$.}
\label{fig:sqd}
\end{figure}

We have tested the above model on $\mu$-SQUID samples. A micrograph of one such device is shown in \mbox{Fig.} \ref{fig:sqd} inset. We use Nb films deposited using DC magnetron sputtering in a chamber with a base pressure in the $10^{-7}$ mbar range. For most of the samples, Nb thin films are deposited on a Si wafer and a photo resist is spun on the films. Using optical lithography, we form a coarse pattern (several  $\mu$m size) on this resist, which is transferred to the film by wet chemical etching using dilute hydrofluoric acid (HF). The final desired pattern is obtained by finer milling with the help of Focused Ion Beam (FIB).

The film thickness $d$ was measured using a profilometer across a step made by masking during deposition. The width $w$ and length $\ell$ of the WLs were estimated from the SEM images. The sample thickness varies between 30 to 75 nm, whereas the width and length of the WLs vary from 50 to 200 nm. All the WL dimensions are thus much smaller than the length $\eta$ defined earlier. For most of our devices, the maximum asymmetry between the two junction is less than 10\%  both in length and width.

In this article, we report on four samples whose detailed parameters are given in Table \ref{tab1}. Transport experiments were performed down to 300 mK in a $^{3}$He cryostat. We used \mbox{r.f.} filters at several stages of the cryostat to minimize noise. Measurements were done in current bias mode using a \mbox{d.c.} current source. No magnetic field was applied for the data presented here.

\begin{table*}
\caption{\label{tab1} Comparison between various experimental (exp) and fit parameters: WL length $\ell$ and width $w$, film thickness $d$, characteristic resistance $R_c$, critical temperature $T_c$, current scale $I_0$.}
\begin{tabular}{|l|c|c|c|c|c|c|c|c|c|}
\hline
Sam.&$\ell$&$w$&$d$&$R_{c}$ (exp)&$R_{c}$ (fit)&$T_{c}$ (exp)&$T_{c}$ (fit)&$I_{0}$ (exp)&$I_{0}$ (fit)\\
no. &(nm)&(nm)&(nm)&$(\Omega)$&$(\Omega)$&(K)&(K)&(mA)&(mA)\\
\hline
1&95&75&45&2.7 $\pm$ 0.3&2.5 $\pm$ 0.5 &5.50 $\pm$ 0.10 &5.25 $\pm$ 0.53&0.64 $\pm$ 0.10&0.81 $\pm$ 0.10\\
2&100&100&55&2.3 $\pm$ 0.2&1.6 $\pm$ 0.3 &5.70 $\pm$ 0.10 &5.60 $\pm$ 0.56&0.77 $\pm$ 0.12&0.78 $\pm$ 0.09\\
3&150&145&65&2.0$\pm$ 0.2&1.9 $\pm$ 0.3 &5.80 $\pm$ 0.10 &5.60 $\pm$ 0.56&0.91 $\pm$ 0.14&0.97 $\pm$ 0.12\\
4&150&150&30&5.3 $\pm$ 0.5&5.1$\pm$ 0.5 &4.5 $\pm$ 0.10 &4.35$\pm$ 0.44&0.27 $\pm$ 0.04&0.24 $\pm$ 0.03\\
\hline
\end{tabular}
\end{table*}

\mbox{Fig.} \ref{fig:sqd} shows the temperature variation of Sample 2 resistance down to 4.2 K. The main and sharp transition with an onset at 7.44 K is expectedly for the bulk film. The other transitions (steps) correspond to relatively smaller pads connected to the WLs. Since the resistance has a large tail, it is difficult to define the critical temperature $T_{c}$ from this data. We therefore define $T_{c}$ from I-V measurements (discussed below) as the temperature above which $I_{c}$ is zero, i.e. the I-V curve is ohmic.

\begin{figure}[tbp]
\centerline{\epsfxsize = 2.35 in \epsfbox{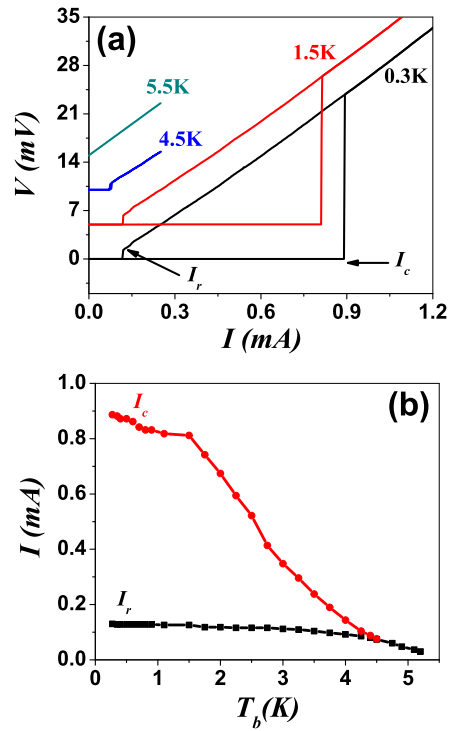}}
\caption{(a) I-V curve of Sample 1 at four different bath temperatures. The plots of 1.5 K, 4.5 K and 5.5 K have been shifted upwards by 5, 10 and 15 mV, respectively, for clarity. (b) Temperature dependence of the critical and retrapping currents for the same sample.}
\label{fig:IV_Tb}
\end{figure}

Sample 1 I-V characteristics at different bath temperatures are shown in \mbox{Fig.} \ref{fig:IV_Tb}a. At low temperature, the curves are clearly hysteretic. From this type of data, we experimentally define the critical current as the maximum current up to which no measurable voltage is observed when the current is ramped from zero. Here, we do not distinguish between critical current and switching current. In the retrapping branch, we define the retrapping current as the current at which the resistance goes back to zero. For most of the samples, with the above definitions, the detection of $I_{c}$ and $I_{r}$ are accurate within about 1\% for $T_{b} < 1$ K and about 10\% near $T_{c}/2$. Close to $T_{c}$, because the transition region width, $I_{c}$ or $I_{r}$ cannot be defined with an accuracy better than 50\%. \mbox{Fig.} \ref{fig:IV_Tb}b shows the variation of critical $I_{c}$ and retrapping $I_{r}$ currents as a function of bath temperature for the same sample. Above a temperature $T_{h}$, the retrapping and the critical currents are equal, meaning that hysteresis in the I-V curve has disappeared.

\begin{figure}[tbp]
\centerline{\epsfxsize = 3.5 in \epsfbox{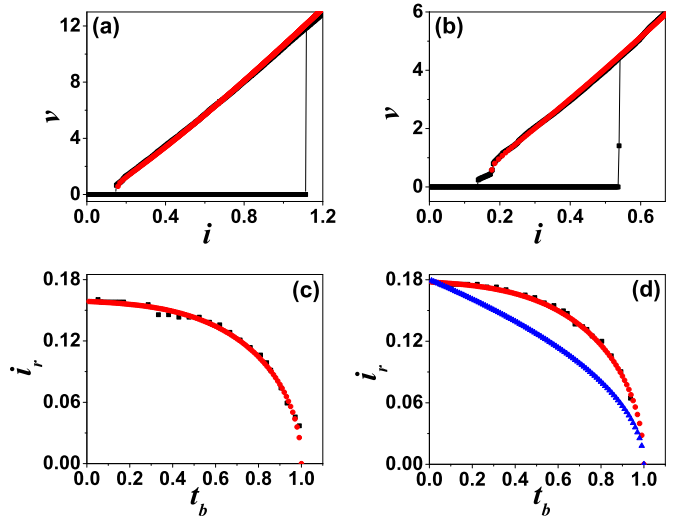}}
\caption{Experimental (black) and numerical fit (red) of normalized I-V curves for Samples 1 and 2 at $T_{b}= 300$ mK with (a) $\beta=3.5$ and $x_{1}=0.015$ for Sample 1 (b) $\beta=3$ and $x_{1}=0.02$ for Sample 2. (c) and (d): Variation of the normalized retrapping current with the normalized bath temperature for Sample 1 and 2. The black dots are the data and the red curves are fits based on \mbox{Eq.} \ref{eq:minima}. The fit parameters are listed in Table \ref{tab1}. The blue curve in (d) is fitting with the Skocpol et al. \cite{tinkham-jap} (their Eq. 14) prediction $i_{r} = 0.18 (1-t)^{1/2}$.}
\label{fig:ir_iv}
\end{figure}

\mbox{Fig.} \ref{fig:ir_iv}a and b show the experimental $I-V$ at the lowest temperature for the two samples together with their fit by our model. The current and voltage are normalized with respect to the fit-derived parameters $I_{0}$ and $R_{c}I_{0}$. In \mbox{Fig.} \ref{fig:ir_iv}c and d, the variation of retrapping current as a function of temperature are shown. Here also the current is normalized with respect to $I_{0}$, whereas the temperature is normalized with respect to the critical temperature $T_{c}$. The values of the fit parameters $R_{c}$, $I_{0}$ and $T_{c}$ together with the experimental parameters are listed in Table \ref{tab1}. Here the experimental $R_{c} = 2\rho_{n}/\pi d$ is calculated by measuring the resistance of a known rectangular geometry and $I_{0} = 2 T_{c}\sqrt{L_{0}}/R_{c}$ is calculated by using the value of above $R_{c}$ and experimental $T_{c}$. We have obtained a similar agreement with Sample 3 and 4 (not shown here), whose experimental and fit parameters are also included in Table \ref{tab1}. We have also used the same model for single WLs and $\mu$-SQUIDs from Hasselbach et al. \cite{hasselbach} with a different geometry. In both cases, we could fit both the low temperature hysteretic I-V curves and temperature dependence of the retrapping current with our model very well.

\mbox{Fig.} \ref{fig:ir_iv}d also shows a fit of the retrapping current with the $\sqrt{1-t}$ dependence from Skocpol et al. \cite{tinkham-jap}. The fitted coefficient 0.18 compares reasonably with the estimated value of 0.34, assuming a Wiedemann-Franz law to get the normal-state thermal conductivity. Nevertheless, our model gives a clearly much better agreement, which we attribute to incorporation of the superconductor thermal conductance temperature dependence in our model.

\section{Discussion}

The exact thermal model for our system is quite involved with complicated non-linear differential equations. In this paper, we have tried to simplify them in a way that the essential features are preserved. This simplified model fits the experimental data very well. Nevertheless, several approximations need further discussion.

We assumed the width of the connecting pads to be much greater than $\eta$, but in actual experiments it is comparable to it. Therefore the actual thermalization would be poorer than what is being assumed; we may be slightly overestimating $x_{1}$. As it is difficult to estimate $\alpha$ for our samples and it actually has a temperature dependence \cite{peroz}, the determination of $\eta$ and hence $x_{1}$ can again be significantly erroneous. For most of the cases, we could fit our data with a 20\% variation in the value of $x_{1}$ by adjusting the other parameters.

We have made the hypothesis that the electron and phonon temperatures are equal in the superconducting region. The electron-phonon coupling power in a volume $V$  is given by \cite{wellstood}, $P =\Sigma V(T_{e}^{5}-T_{p}^{5})$, where $\Sigma$ = $2.4 \times 10^{9}$ W.m$^{-3}$K$^{-5}$ is the electron-phonon coupling parameter for Nb. Taking typical experimental values $I_{r}$ = 0.12 mA, $R_{c}$ = 2.4 $\Omega$, $r_{0}/r_{1} = 1.1$ (that gives $R(r_{0})$ = 8 $\Omega$), we get a dissipated power P = $I_{r}^{2} R(r_{0})$ = 0.12 $\mu$W. Nearly all of this resistive heat is transmitted to the substrate in the superconducting region only. Though the temperature decreases sharply making the heat loss rather non-uniform, the effective size of this region is of order $\eta$, which ranges between 1 and 3 $\mu$m. Taking $\eta$ = 2 $\mu$m, d = 50 nm and an average electron temperature $T_{e}= 4$K, the volume of the superconducting region is, $V=\pi\eta^{2} d$ = $3.5 \times 10^{-19} m ^{3}$. This gives the temperature difference $T_{e}-T_{p}$ as 0.06 K only, validates our hypothesis.

We took a linear approximation for the surface loss term. The metal film and the substrate phonons exchange heat through a Kapitza resistance, giving a power $K_a(T^{4}-T_{b}^{4})$ per surface unit, with $K_a$ as the Kapitza constant \cite{wellstood}. For $T$ close to $T_b$, the above expression can be approximated as $4K_aT_b^3(T-T_b)$. From the temperature profile in Fig.\ref{fig:temp_dist}, we can say that for most of superconducting region the above approximation is valid except for the region close to the N-S interface ($r \simeq r_0$). However, if the bath temperature is close to the critical temperature then for the entire WL the above approximation would be valid.

We also neglected the heat loss to the substrate from the normal region, \mbox{i.e.} the WL and the semicircular region between $r_1$ and $r_0$. Let us compare the heat transfer to the substrate $P_{s}$ and the heat conducted out $P_{c}$ under the linear approximation. Considering only one half of the film, we can approximately write $P_{s} = \alpha \pi r_{0}^{2}(t_{1}-t_{b})T_{c}/2$. We can also write $P_{c} = -\kappa_{n}\pi r_{0}d(\frac{dT}{dr})_{r = r_{0}} = \pi r_{0}d L_{0}T_{c}^2(t_{1}-1)/(\rho_{n}(r_{0}-r_{1}))$. Here we assume a linear temperature decrease within the normal region, which is fairly justified according to \mbox{Fig.} \ref{fig:temp_dist}c, d. For $t_{1}= 1.1$, $t_{b}= 0.05$, $r_{1}=100$ nm, $r_{0} = 2r_{1}$, $\alpha$ = 5 W/cm$^{2}$.K, $d = 50$ nm, $\rho_{n}$ = 25 $\mu\Omega$ cm, and $T_{c} = 8$ K, one gets $P_{c}$ = 0.2 $\mu$W and $P_{s}/P_{c} \simeq 0.1$, which confirms our assumption.

The surface heat loss from the WL normal state region was neglected by assuming $(t_{1}-t_{b})\frac{A}{2\pi\eta^2}\ll\beta i^{2}$. Let us check the argument for the worst possible case; \mbox{i.e.} at lowest possible temperature and for longer WLs. Taking  $t_{b} = 0.05$, $t_{1} = 1.1$, $A=$ 300 $\times$ 300 nm$^2$, $\eta$ = 1 $\mu$m at $i=i_{r} = 0.15$, one gets $(t_{1}-t_{b})\frac{A}{2\pi\eta^2} = 0.015$, whereas with $\beta = \pi$ (which corresponds to $\ell = w = 300$ nm) and $i = i_{r} = 0.15$ we get $\beta i^{2} = 0.072$, which is almost 5 times higher than $(t_{1}-t_{b})$$\frac{A}{2\pi\eta^2}$.

\section{When does hysteresis disappear?}

\begin{figure}[tbp]
\centerline{\epsfxsize = 2.25 in \epsfbox{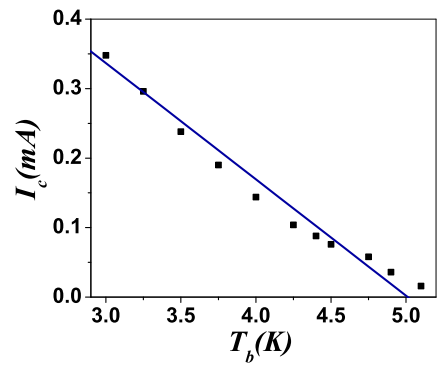}}
\caption{Variation of critical current with the bath temperature in the high temperature regime for Sample 1. The blue line is a straight line fitting $I_{c}=I_{c0}(1-T_{b}/T_{c}$) with $I_{c0}= 0.84$ mA and $T_{c}= 5.05$ K.}
\label{fig:Ic}
\end{figure}

A key feature is the disappearance of hysteresis at high temperature. In general, there is a particular bath temperature $T_{h}$ at which $I_{r}$ and $I_{c}$ are equal. Above this temperature, for $I>I_c$ the current is large enough to kill the superconductivity in the WL, making it resistive. But the related Joule heating is not sufficient to raise the WL temperature above $T_c$ and provide an N-S interface. In order to find the crossover temperature $T_{h}$, we need an expression for the temperature dependence of the critical current. For a bath temperature near the critical temperature, we can use the expression: \cite{likharev,likharev1} $I_{c}R_{n}= \gamma T_{c}(1-t_{b})$, where $\gamma$ = 635 $\mu$V/K and $R_{n}$ is the normal state resistance. In practice, $\gamma$ can vary significantly. Taking $R_{n}$ = $R_{0}$, one can simplify the above equation to:
\begin{eqnarray}
i_{c}= \frac{\gamma}{2\beta\sqrt{L_0}}(1-t_b).
\label{eq:ic}
\end{eqnarray}
Here, $i_{c}=I_{c}/I_{0}$ and we have used $\gamma = 635 \mu$V/K, $L_0 = 2.44\times10^{-8} $W.$\Omega$/K$^2$. In \mbox{Fig.} \ref{fig:Ic}, we plot the variation of $I_{c}$ with the bath temperature for Sample 1 above $T_{c}/2$. From the linear fit, we extract $\gamma = 930 \mu$V/K with the above $L_0$ value, $\beta = 3.5$ and $R_{c} = 2.28 \Omega$.

\begin{figure}[tbp]
\centerline{\epsfxsize = 2.35 in \epsfbox{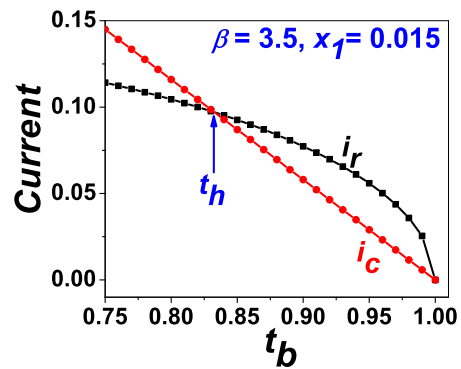}}
\caption{Temperature dependence of $i_{c}$ (red) following \mbox{Eq.} \ref{eq:ic} and of $i_{r}$ (black) calculated from our model, near the critical temperature. The parameters used for the plot are given in the figure. At $t_{b}=t_{h}$, the two curves cross each other, so that hysteresis disappears for higher temperatures.}
\label{fig:ir_ic}
\end{figure}

In \mbox{Fig.} \ref{fig:ir_ic}, we plot the variation of $i_{c}$ (red curve) and $i_{r}$ (black curve) as a function of the normalized bath temperature $t_b$ near $t_b=1$ using \mbox{Eq.} \ref{eq:ic}, for $\beta=3.5$ and $x_{1}= 0.015$. The crossover temperature $t_h$ is then straightforwardly determined from the intercept of the two curves. Let us point out here that the critical current $I_c$ and the retrapping current $I_r$ are controlled by two different physics, with $I_c$ dependent on the WL superconducting properties and $I_r$ on the heat dissipation. This justifies that these currents have a different temperature dependence.

\begin{figure}[tbp]
\centerline{\epsfxsize = 2.5 in \epsfbox{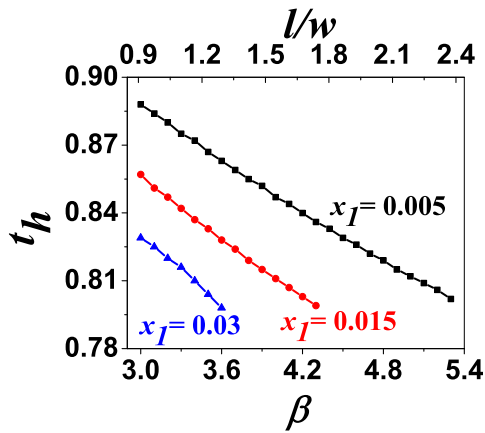}}
\caption{Variation of the hysteresis crossover temperature $t_{h}$ with the parameter $\beta$ for three different values of $x_{1}$ (\mbox{i.e.} WL width). The top axis represents $\ell/w$, calculated using the formula: $\beta$= $\frac{\pi}{2}$$(1+\frac{\ell}{w})$.}
\label{fig:ir_ic2}
\end{figure}

In \mbox{Fig.} \ref{fig:ir_ic2}, we plot the variation of $t_{h}$ as a function of $\beta$ for three different values of $x_{1}$. The top axis refers to $\ell/w$, which is related to $\beta$ by the formula: $\beta$ = $\frac{\pi}{2}(1+\frac{\ell}{w})$. The upper limits $\beta_{max}$ of the parameter $\beta$ are chosen in a way that at this point $x_{0}$ = $x_{1}$, \mbox{i.e.} the N-S interface is at the WL boundary, as beyond this value our short WL approximation does not hold. From \mbox{Fig.} \ref{fig:ir_ic2}, this occurs at a $t_{h}$ of about 0.8.

\begin{figure}[tbp]
\centerline{\epsfxsize = 2.5 in \epsfbox{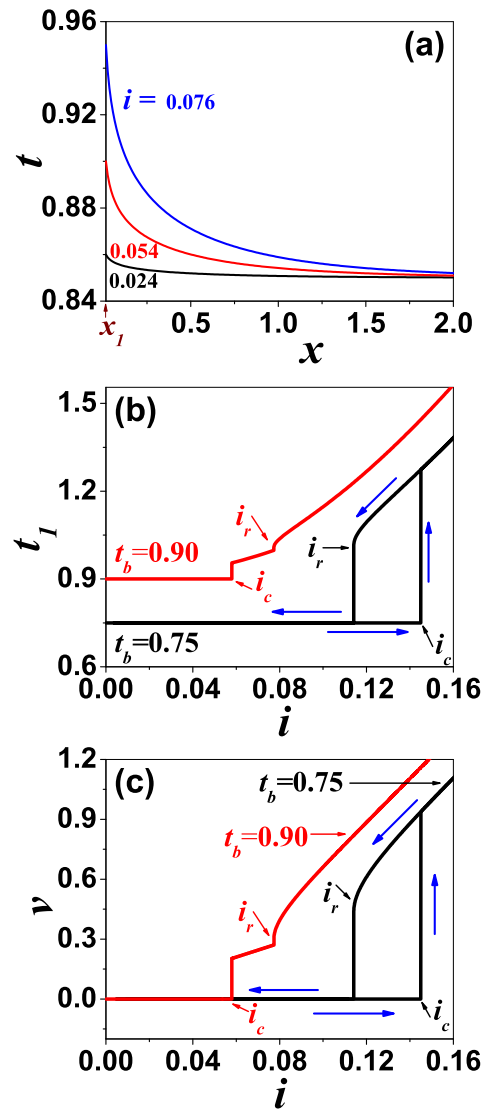}}
\caption{Calculation results with parameters $\beta$ and $x_{1}$ equal to 3.5 and 0.015 respectively, which give $t_{h}$= 0.83. (a) Calculated temperature distribution at a bath temperature $t_{b}$ = 0.85 above the threshold temperature $t_{h}$ for normalized bias currents of 0.024, 0.054 and 0.076. Here $i_{r}$ is 0.093. (b,c) Variation of WL temperature (b) and $i-v$ curve (c) as a function of bias current for bath temperatures above the threshold $t_{b} = 0.75$ (black) and below $t_{b}= 0.9$ (red). The blue arrows indicate the direction of current sweep.}
\label{fig:ir_ic1}
\end{figure}

In the hysteresis-free regime $t>t_h$, the detailed temperature profile $i_r>i>i_c$ can be found by solving \mbox{Eq.} \ref{eq:hf7} for $x>x_1$, \mbox{\mbox{i.e.}} the superconducting region. The boundary conditions used for solving this equation are 1) at $x=x_1$, -$x_1t_1\frac{dt}{dx}=\frac{\pi d R_0i^2}{2\rho_N}$ and 2) $t=t_b$ as $x\rightarrow\infty$. Solutions for $t$ were found numerically and are plotted in \mbox{Fig.} \ref{fig:ir_ic1}a.

\mbox{Fig.} \ref{fig:ir_ic1}b shows the temperature $t_{1}$ of the WL as a function of current for two bath temperatures, $t_b=0.90$ and $t_b=0.75$, respectively above and below the hysteresis threshold $t_h=0.83$. As the current is ramped up from zero, the WL temperature jumps from $t_1=t_b$ to a higher value at $i=i_c$. For a bath temperature above the hysteresis threshold ($t_b>t_h$), there is another jump in WL temperature at $i=i_r$ and after this the temperature keeps on increasing. The $t_1$ vs $i$ behavior remains same when $i$ is ramped down, \mbox{i.e.} there is no hysteresis. Let us point out that there is no actual retrapping at this $i_r$ value, but the appearance or disappearance of a N-S interface close to the WL. For a bath temperature below the hysteresis threshold ($t_b<t_{h}$), the behavior shows an upward jump from $t_1=t_b$ to a higher value when the current is ramped up through $i=i_c$, and a downward jump to $t_1=t_b$ when current is ramped down through $i=i_r$. Hysteresis is thus present.

\mbox{Fig.} \ref{fig:ir_ic1}c shows the $i-v$ curves as calculated from the location of the N-S interface (above $i_r$) and the resistance of the WL (below $i_r$) at the same two bath temperatures above and below $t_h$. We see a close resemblance between the temperature and voltage curves as a function of bias current. While below $t_h$, the $i-v$ spectra describe well the experimental curve, above $t_h$ the calculated curve shows an extra step at $i_r$ arising from the sudden creation of the N-S interface at a position $r_0>r_1$. This step was not observed in experiments. We believe that the predicted extra step may get significantly rounded as the WL temperature approaches $T_c$. The superconducting region outside $r_1$ will be close to $T_c$, reducing its critical current density. The exact shape of the I-V curve in the non-hysteretic regime will then be dictated by thermally activated phase slips \cite{bezryadin,skocpol-tinkham} for $i<i_c$ and superconducting fluctuations for $i>i_c$ \cite{skocpol-tinkham}.

Let us now consider how one could manipulate the crossover temperature $t_h$. At a fixed $x_{1}$ (which is proportional to the width $w$), $t_{h}$ decreases with the increase of $\beta$ (and hence the length $\ell$), see \mbox{Fig.} \ref{fig:ir_ic2}. This is desirable if we want hysteresis to disappear at low temperature. But the adjustment of $\beta$ to any arbitrary value is impossible, since we wish the WL to behave like a Josephson Junction, which implies the condition: w $\leq$ $\ell$ $\sim$ $\xi$ \cite{likharev}. This gives a lower bound on $\beta$ and $x_{1}$ and hence $t_{h}$ in general. Therefore, it is generally not possible to eliminate hysteresis for these WL junctions just by manipulating the WL width and length.

However, since $\beta$ is the ratio of WL resistance $R_0$ to the characteristic resistance $R_c$ of the film, we can effectively increase $\beta$ by increasing the WL resistance. This can be done by reducing the thickness of the WL alone. The reduction of the whole film thickness ($d$) including the connecting electrodes can also reduce $T_h$: in this case, $\eta\propto\sqrt{d}$ is smaller, giving a larger $x_0$. This makes hysteresis disappear at smaller temperatures (see \mbox{Fig.} \ref{fig:ir_ic2}). If the critical temperature $T_{c}$ is not affected, a wider temperature span for the hysteresis-free regime is achieved. This improvement was observed by Tinkham et al. \cite{tinkham-prb} for superconducting nano-wires. With an appropriate choice of substrate and growth conditions, the resistivity and critical temperature of very thin films can remain almost unaffected by thickness reduction (for Nb see Ref. \onlinecite{bouchiat}), enabling similar results to be obtained with continuous films. Another possible way is to reduce the critical current $I_c$ (and possibly increase $\beta$) by either making the WL with a poor superconductor or completely replacing it by a normal metal. Angers et al. \cite{angers} were able to get a $T_h<1.2 K$ for $\mu$-SQUIDs made with SNS type weak links based on Nb.

\section{Conclusions}

In conclusion, we have described a thermal model for understanding the hysteresis in the I-V curve of short superconducting WLs and their extension to $\mu$-SQUIDs. Using this model, we have calculated the detailed I-V characteristics and the temperature profile near the WL as a function of bath temperature. We have obtained a good agreement between experiments and theory in terms of I-V characteristics and their temperature dependence. A key finding of this model, which again is in agreement with the experiments, is the disappearance of hysteresis above certain temperature. We have discussed how one can adjust the WL geometry in order to widen the temperature range of this hysteresis-free regime of WL-based $\mu$-SQUIDs.

\section{Acknowledgements}

DH acknowledges the financial support from CSIR, India and the French Embassy in India. Discussions with Klaus Hasselbach have been fruitful. Sincere thanks to Sudhanshu Srivastava for help with FIB, Prabhat Kumar Dwivedi for his help in optical lithography. Thanks to Franck Dahlem, Thomas Quaglio and Soumen Mandal for helping in some of the technical matters. LP acknowledges financial support from R\'egion Rh\^one-Alpes. AKG thanks Universit$\acute{e}$ Joseph Fourier, Grenoble for its support during his visit in 2009.


\begin{thebibliography}{99}
\bibitem{likharev1} Here, instead of $\gamma$, Likharev \cite{likharev} used the symbol $\alpha$. Since we have already used $\alpha$ for surface heat loss coefficient, here we use a different symbol.
\end{thebibliography}
\end{document}